\documentclass[%
 reprint,
 amsmath,amssymb,
 aps
]{revtex4-2}

\usepackage{graphicx}
\usepackage{dcolumn}
\usepackage{bm}
\usepackage{amsmath}
\usepackage{graphicx}
\usepackage{array}
\usepackage{booktabs}
\usepackage{threeparttable}
\usepackage[hypertexnames=false]{hyperref}
\usepackage{xcolor} 
\definecolor{darkblue}{RGB}{0,0,139}
\hypersetup{
    colorlinks=true,                     
    linkcolor=darkblue,                  
    urlcolor=darkblue,                   
    citecolor=darkblue,                  
    filecolor=darkblue,                  
    bookmarksnumbered=true               
}
\usepackage{ulem}
\usepackage{titlesec}
\usepackage{tabularray}  
\usepackage{multirow}   
\usepackage{amssymb}    
\usepackage{hhline}
\usepackage{makecell}
\usepackage{array}
\usepackage{xpatch}   

\renewcommand{\emph}[1]{\textit{#1}} 

\renewcommand{\thetable}{\arabic{table}} 
\renewcommand{\theequation}{\arabic{equation}}  
\begin{document}

\preprint{APS/123-QED}

\title{
Nonlinear Maxwell's Equations as a Boundary Value Problem in Multilayer Nanophotonics
}

\author{Ruizhe Gu$^{1}$, Dingyang Zhang$^{1}$, and Wenjing Liu$^{1,2,\dagger}$\\
\vspace{6pt}
$^1$State Key Laboratory for Artificial Microstructure and Mesoscopic Physics and Frontiers Science Center for Nano-optoelectronics, School of Physics, Peking University, Beijing, 100871, China\\
$^2$Collaborative Innovation Center of Extreme Optics, Shanxi University, Taiyuan, 030006, China\\
$^{\dagger}$Corresponding author: wenjingl@pku.edu.cn}
\date{\today}

\begin{abstract}
Calculating strong‑field, broadband nonlinear phenomena in complex nanophotonic structures remains difficult because it requires simultaneous handling of non-perturbative dynamics, impulsive processes, and structural complexity.
Here, we introduce a unified framework that combines the transfer matrix method with an iterative Green’s function approach. 
The linear multilayer response is incorporated as boundary conditions into a nonlinear boundary value problem, which is then solved self-consistently, naturally accommodating broad spectra and arbitrary nonlinearity.
The framework is validated with two examples in the transparent and resonant regimes, respectively.
The first demonstrates an impulsive spectral modulation in second harmonic generation that is captured only by a full-structure broadband treatment.
The second demonstrates the entire dynamical evolution of exciton-polaritons in a pump-probe experiment reproduced by a unified simulation. 
This framework directly links nanophotonic design with ultrafast nonlinear dynamics.

\end{abstract}

\maketitle

\textit{Introduction.}- Nonlinear optical effects lie at the core of cutting-edge optical research both fundamentally and technologically \cite{boyd2020nonlinear}. 
At the same time, the push toward integrated nanophotonics requires on-chip implementation and precise characterization of nonlinear processes \cite{chen2012nonlinear, li2017nonlinear, panoiu2023fundamentals, dutt2024nonlinear}. 
When mediated by nanophotonic structures, nonlinear processes can exhibit several distinctive features.
First, strong field confinement in wavelength-scale structures drives nonlinear processes such as self-modulation and pump depletion into a non-perturbative regime even at modest input powers.
Second, ultrafast laser pulses have been widely applied due to short interaction lengths, which introduce rich impulsive interactions.
Third, resonant nanostructures can support strongly enhanced light-matter coupling that substantially shapes the light fields through the material dynamics, which demands a self-consistent treatment.

Simulating these processes therefore requires a method that handles simultaneously non-perturbative processes, broadband spectrum ranges, and complex multilayer geometries \cite{Lavrinenko2014}.
In the time domain, the finite-difference time-domain (FDTD) method becomes computationally demanding, as covering wide spectra requires ultrasmall time steps and incorporating dispersive media requires intricate material models \cite{ziolkowski1997incorporation, shokooh2005analysis}. 
The beam propagation method, on the other hand, struggles with the multiple reflections and large index contrasts inherent to multilayer geometries \cite{scalora1994beam}.
In contrast, frequency domain methods, such as the transfer matrix method (TMM), hold promises to overcome these limitations, as they naturally handle broadband responses with complex structures in linear problems.
Extending these methods into the nonlinear regime has therefore been a long-standing goal, which may offer a powerful tool for nonlinear thin-film devices. 

However, most existing works adopted approximations to varying degrees, which limits their application scenarios. 
Small-signal approximation or linearization methods were widely applied to formulate various parametric processes with TMM \cite{bethune1989optical, hashizume1995optical, li2007second, ren2010enhanced, pakhomov2021modeling, ghaemi2022asymmetric, poveda2025parametric}, which are not suitable for strong field applications.
Some works applied iteration methods to obtain self-consistent solutions without linearization, but only considered continuous or quasi-monochromatic light fields \cite{maes2004modeling, saleh2008second, xia2009enhancement, ren2011analysis, yuan2017continuation, amotchkina2017experimental, huang2018new}.
Recently, bidirectional pulse propagation method was developed to incorporate ultrashort pulses and dispersive nonlinearity in single slab geometries \cite{kolesik2012quantifying, jakobsen2014bidirectional, hofstrand2019bidirectional, jakobsen2025general}, but its applications in multilayer structures remain untested.
Overall, a unified framework that properly handles ultrafast laser pulses, arbitrary nonlinearities, and complex nanophotonic structures is needed.

Here, we present a unified and rigorous framework for nonlinear multilayer optical systems, by integrating the TMM and iterative Green's function method, where TMM was applied to project the linear responses as boundary conditions of the nonlinear Maxwell's equations, which was subsequently solved by the Green's function.
Two representative nonlinear processes were systematically studied utilizing this framework in non-resonant and resonant frequency ranges of the nonlinear materials, respectively,  where both the optical geometry and material responses  are incorporated simultaneously.

\begin{figure}[htb]
\includegraphics{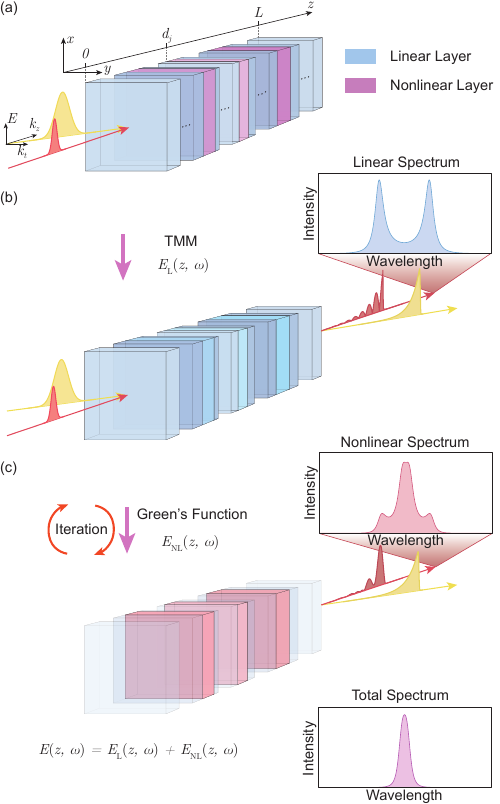}
\caption{\label{fig:schematic} 
\textbf{Schematic of the boundary-projected framework for solving nonlinear Maxwell's equations.}
The electric field is divided into the linear ($E_{\mathrm{L}}$) and nonlinear ($E_{\mathrm{NL}}$) parts and solved sequentially. 
\textbf{a,} The 1D multilayer nonlinear optical structure under external excitations.
\textbf{b,} The linear solution $E_{\mathrm{L}}$ solved by the TMM.
\textbf{c,} The nonlinear boundary value problem obtained by projecting $E_{\mathrm{L}}$ as the boundary conditions, and solved by the iterative Green's function approach.
}
\end{figure}

\textit{Algorithm.}- We consider the solution of nonlinear Maxwell's equations in a one-dimensional (1D) multilayer optical structure under external excitations. 
The structure contains an arbitrary sequence of linear and nonlinear layers stacked along the $z$ axis, with each layer homogeneous in the transverse $x\textrm{-}y$ plane, as shown in Fig. \ref{fig:schematic}(a). 
In nonlinear layers, nonlinear polarization $\vec{P}_{\mathrm{NL}}$ is introduced to the electric displacement vector $\vec{D}$ through $ \vec{D} = \varepsilon_0 n^2\vec{E}+\vec{P}_{\mathrm{NL}}$, where $\varepsilon_0$ is the vacuum permittivity, $n$ is the linear refractive index, and $\vec{E}$ is the electric field.
Owing to the translational invariance in both temporal and transverse spatial dimensions, the electric field can be expanded by $\vec{E}(\vec{k}_t, z, \omega)$, with $\vec{k}_t$ and $\omega$ denoting the transverse momentum and the angular frequency, respectively. 
The field is further divided into transverse-electric (TE) and transverse-magnetic (TM) polarizations, with the electric field perpendicular and parallel to the incident plane, respectively, which results in scalar solutions for each polarization. 
Without loss of generality, only the TE polarization is discussed in the main text, and readers are referred to Appendix A for a complete discussion of the TM polarization.

The wave equation with nonlinear polarization and boundary conditions can be written as
\begin{subequations}\label{eqset:WaveEqn}
    \begin{gather}
    \left(\frac{\partial}{\partial z^2} + \frac{n^2\omega^2}{c^2}-k_t^2\right)E(z) = -\mu_0\omega^2 P_{\mathrm{NL}}(z), \label{eq:NLwave}\\
    E|_{z=d_{j,-}}=E|_{z=d_{j,+}}, \quad \left.\frac{\partial E}{\partial z}\right|_{z=d_{j,-}}=\left.\frac{\partial E}{\partial z}\right|_{z=d_{j,+}},\label{eq:BCLayer} \\
    \frac{1}{2}\left.\left(E+\frac{\mathrm{i}}{k_z}\frac{\partial E}{\partial z}\right)\right|_{z=0_-}=E_{\mathrm{inc}}, \label{eq:BCInc} \\
    \frac{1}{2}\left.\left(E-\frac{\mathrm{i}}{k_z}\frac{\partial E}{\partial z}\right)\right|_{z=L_+}=0, \label{eq:BCTrans}
    \end{gather}
\end{subequations}
where $\mu_0$ is the vacuum permeability, $c$ is the vacuum speed of light, $k_z=\sqrt{n^2\omega^2/c^2 - k_t^2}$ is the wavenumber along the $z$ axis, $0, d_j, L$ are the positions of the top surface, the $j^{\mathrm{th}}$ interface, and the bottom surface, respectively.
$E_{\mathrm{inc}}$ is the amplitude of the incident electric fields.
Mathematically, Eq. \eqref{eqset:WaveEqn} represents a nonlinear 1D boundary value problem (BVP) with both nonhomogeneous excitations and boundary conditions, induced by nonlinear polarizations and linear excitations, respectively. 
Thus, the electric field can be divided into 
\begin{equation}
    E(z)=E_{\mathrm{L}}(z) + E_{\mathrm{NL}}(z).
\end{equation}
where $E_{\mathrm{L}}$ is the solution of the linear wave equation under nonhomogeneous boundary conditions and $E_{\mathrm{NL}}$ is the solution of the nonhomogeneous wave equation under homogeneous boundary conditions, as illustrated in Fig. \ref{fig:schematic}(b) and (c).
Physically, $E_{\mathrm{L}}$ is the linear electric field distribution built up by external excitations and can be solved by TMM directly.
$E_{\mathrm{NL}}$ is the distribution generated by nonlinear polarization and shaped by the multilayer structure.
To solve $E_{\mathrm{NL}}$, the linear layers are first projected as homogeneous boundary conditions of the nonlinear layers by TMM (Fig. \ref{fig:schematic}(c)).
The nonlinear wave equation is then solved by the Green's function method iteratively. 
Detailed description of the algorithm is presented in Appendix A.
In the following, we investigate typical nonlinear processes based on practical experimental schemes.

\begin{figure*}[htb]
\includegraphics{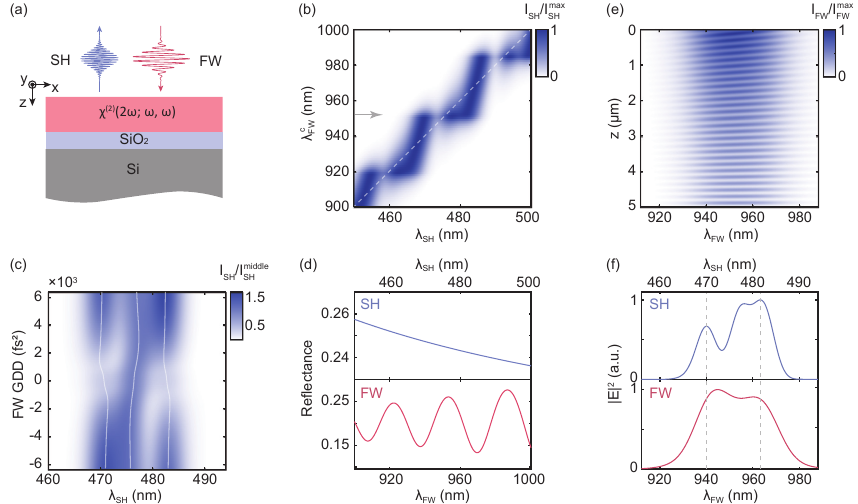}
\caption{\label{fig:SHG} {\bfseries Impulsive spectral modulation of second harmonic generation.}
\textbf{a,} Schematic of the SHG simulation setup. A thin film with $\chi^{(2)}$ is placed on the $\mathrm{SiO_2/Si}$ substrate. 
The FW pulse is normally incident onto the surface, generating a reflected SH pulse.
\textbf{b,} SH spectrum as a function of the FW center wavelength.
The duration and linewidth of the FW pulse are set as 200 fs and 30 nm, respectively, with positive GDD.
\textbf{c,} SH spectrum as a function of the GDD of the FW pulse at $\lambda_{\mathrm{FW}}^c$ = 952 nm, as indicated by the gray arrow in \textbf{b}.
White lines indicate the peak positions extracted from Gaussian fitting of each peak.
Each spectrum is normalized to the peak amplitude of the middle peak.
\textbf{d,} Reflectance spectra of the multilayer structure in the FW band (red line) and the SH band (blue line).
\textbf{e,} FW field distribution inside the nonlinear layer, with $\lambda^c_{\mathrm{FW}}$ = 952 nm.
\textbf{f,} Comparison between the SH spectrum and the FW spectrum at the top interface of the nonlinear material ($z=0$).
Gray dashed lines indicate the peak wavelengths of the two side peaks of the SH. 
}
\end{figure*}

\textit{Impulsive spectral modulation for second harmonic generation.}- Second harmonic generation (SHG) is a nonlinear optical process frequently observed in non-centrosymmetric crystals.
The nonlinear polarization can be defined as $P_{\mathrm{NL}}(z,t) = \varepsilon_0\chi^{(2)}(z)E^2(z,t)$, where two fundamental wave (FW) photons interact through second-order nonlinear susceptibility $\chi^{(2)}$ to generate a second harmonic (SH) photon.
Femto- and pico-second lasers are widely applied in nanoscale SHG studies to obtain strong, ultrafast excitations within short interaction lengths.
SHG spectroscopy provides invaluable insights in structural symmetry and nonlinear susceptibility of materials \cite{shen1989surface, Fiebig2005, Denev2011, Wang2019, sun2019giant}, as well as exotic light-matter interactions such as polariton Fano resonances \cite{wang2017exciton} and quantum interferences \cite{lin2019quantum, qian2024probing}.
However, as will be shown in this section, to accurately interpret the spectral features of SHG, it is necessary to carefully analyze impulsive processes that are strongly modulated by the multilayer geometry.

The structure in the calculation consists of a 5 $\mathrm{\mu m}$ thick nonlinear film on top of a 277 nm $\mathrm{SiO_2}$/Si substrate, a typical geometry in studies of two-dimensional nonlinear materials (Fig. \ref{fig:SHG}(a)).
To be consistent with realistic experimental parameters, here we applied susceptibilities of $\mathrm{NbOI_2}$, a two-dimensional ferroelectric material possessing strong second order nonlinearity \cite{Fang2021, abdelwahab2022giant, Fu2024}. 
A FW pulse with a Gaussian profile of 30 nm linewidth is normally incident on the structure.
The reflected SH spectrum as a function of the center wavelength of the FW $\lambda^{c}_{\mathrm{FW}}$ is shown in Fig. \ref{fig:SHG}(b), with each row normalized to its maximum.
Interestingly, SH signals periodically deviate from $2\lambda^c_{\mathrm{FW}}$, accompanied by a spectral splitting into multiple peaks.
These multi-peak features are verified to originate from impulsive processes by scanning the group delay dispersion (GDD) $\phi_2$ of the FW pulse while fixing its spectrum centered at $\lambda^c_{\mathrm{FW}} = 952$ nm, as shown in Fig. \ref{fig:SHG}(c).
At this $\lambda^c_{\mathrm{FW}}$, the SH spectrum splits into three peaks with both wavelengths and intensities varying with $\phi_2$.
The middle peak, specifically, weakens as $|\phi_2|$ increases, while its peak wavelength exhibits a kink opposite to the two side peaks as $\phi_2$ changes from negative to positive.

To understand the spectral characteristic of SHG, the linear reflectance was calculated at the FW and SH wavelengths in Fig. \ref{fig:SHG}(d).
$\mathrm{NbOI_2}$ has a bandgap of 2.24 eV \cite{abdelwahab2022giant}, which lies between the SH ($\sim$ 2.6 eV) and FW ($\sim$ 1.3 eV) photon energies.
Therefore, the reflectance spectrum in the SH band is flat as a result of strong absorption, which oscillates periodically in the FW spectral range due to thin film interference (Fig. \ref{fig:SHG} (d), see refractive index of $\mathrm{NbOI_2}$ in Supplemental Material section S1 A).
By comparing Fig. \ref{fig:SHG}(b) and (d), one can see that the period of the spectral modulation of SHG matches the oscillation period of the reflectance in the FW spectral range, where the splitting of the SH spectrum occurs at the peak of the FW reflectance, highlighting the vital role played by the geometry. 
To further understand this correlation, the electric field distribution of the FW inside the nonlinear layer is calculated with $\lambda^c_{\mathrm{FW}}=952$ nm, as shown in Fig. \ref{fig:SHG}(e).
At 952 nm, the round-trip phase of the FW is odd multiples of $\pi$, leading to destructive interference on the sample surface at $z=0$. 
Away from this wavelength, the round-trip phase evolves toward even multiples of $\pi$, resulting in gradual enhancement of the FW field. 
The large spectral span of the femtosecond light therefore gives rise to a double-peaked spectrum at $z=0$ (Fig. \ref{fig:SHG}(f), lower panel).

\begin{figure*}[htb]
    \centering
    \includegraphics{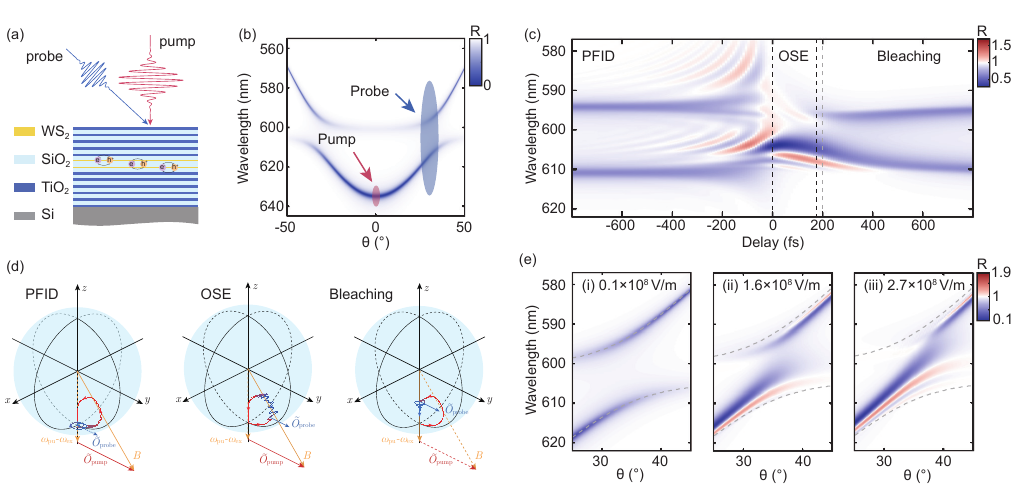}
    \caption{\label{fig:Bleaching} {\bfseries Nonlinear dynamics of exciton-polaritons.}
    \textbf{a,} Schematic of the microcavity exciton-polariton system and the pump-probe setup.
    Three $\mathrm{WS_2}$ monolayers with 6 nm $\mathrm{SiO_2}$ spacer between each layer are embedded a DBR microcavity, excited by the pump and probe pulses.
    \textbf{b,} Linear angle-resolved reflectance spectrum of the microcavity exciton-polaritons calculated by the TMM.
    The wavelength and incident angle of the pump and probe pulses are indicated by the red and blue arrows, respectively. 
    \textbf{c,} Time-resolved reflectance under the pump field amplitude of $1.6\times 10^8$ V/m as a function of the pump-probe delay. 
    The PFID, OSE, and bleaching dynamic stages are divided by the dashed black lines.
    \textbf{d,} Schematics of the exciton dynamics under the three stages of the pump-probe delay.
    The trajectory of the Bloch vector represented by the blue and red curves is visualized in the Bloch sphere, with the arrows indicating the temporal direction of evolution. The red and blue parts are driven by the pump and probe pulses, respectively.
    \textbf{e,} Angle-resolved reflectance at +200 fs pump-probe delay, as indicated by the gray dashed in in \textbf{c}, under the pump field amplitude of (i) $0.1\times 10^8$ V/m, (ii) $1.6\times 10^8$ V/m, and (iii) $2.7\times 10^8$ V/m.
    }
\end{figure*}

During the SHG process, this spectrum is imprinted onto the SH field distribution through nonlinear polarization $P_{\mathrm{NL}}(z)$ (see Supplemental Material section S1 B).
As a result, the reflected SH spectrum (upper panel of Fig. \ref{fig:SHG}(f)) inherits the two side peaks from the FW spectrum indicated by the two gray dashed lines.
The additional central peak is therefore identified as impulsive SHG, as its intensity depends strongly on the FW GDD (Fig. \ref{fig:SHG}(c)). 
The dependence of the SH spectrum on the GDD of the FW pulse can be attributed to the thin film interference induced oscillating spectral phase of the FW field.
This phase modulation breaks the time reversal symmetry between the positively and negatively chirped FW field, and hence introducing spectral modulation in the impulsive SHG processes (see Supplemental Material section S1 C for detailed discussion).
These results demonstrate the ability of our method to calculate broadband, impulsive nonlinear processes, which is important for interpreting nonlinear spectroscopy.
The framework can be readily applied to higher order nonlinearities, and an example on the optical Kerr effect is presented in the Supplemental Material section S2. 

\textit{Nonlinear dynamics of exciton-polaritons.}- While nonlinear materials are favorable for developing high energy and low loss nonlinear devices in their transparency windows, they generally support rich and non-perturbative nonlinear phenomena under resonant excitations.
Understanding and manipulating these phenomena, especially within tailored photonic environments, is critical for cavity quantum electrodynamics and active photonic devices. 
A typical system under investigation is cavity exciton-polaritons. 
Excitons are bosons formed by electron-hole pairs at the band edge of semiconductors, which can strongly couple to cavity photons to form quasi-particles called exciton-polaritons \cite{weisbuch1992observation, deng2010exciton}.
Combining strong nonlinearity of excitons and high degrees of freedom of optical structures, polaritons have demonstrated ultrafast modulations of photonic dispersion \cite{Pickup2020, lamountain2021valley, Chen2022, zhao2023exciton, del2024non} and reconfigurable on-chip devices \cite{zasedatelev2019room, chen2022optically, Lee2024, zhao2025sub, Tassan2026}. 
However, simulations on these systems generally rely on the semi-phenomenological Gross-Pitaevskii equations and are limited to specified nonlinear processes. 
Here, we use our framework to perform a rigorous simulation of nonlinear exciton-polaritons in a planar microcavity based on the optical Bloch equations, providing a unified description of three nonlinear processes under distinct pump-probe stages.

The exciton-polariton system is shown in Fig. \ref{fig:Bleaching}(a), which consists of 3 monolayers of $\mathrm{WS_2}$ embedded in a distributed Bragg reflector (DBR) microcavity (see Supplemental Material section S3 A for detailed parameters, section S3 B for derivation and implementation of the optical Bloch equations). 
The linear reflectance of the coupled system was first calculated by a linear TMM, with the formation of exciton polaritons characterized by an anti-crossing dispersion presented in Fig. \ref{fig:Bleaching}(b).
Then, the nonlinear dynamics of the system was investigated in a pump-probe scheme, with the normal-incident pump pulse centered at the lower polariton $\omega_{\mathrm{pu}}$, and the broadband probe pulse centered at the exciton frequency $\omega_{\mathrm{ex}}$, where both pulses have a temporal duration of 100 fs. 
The transient reflectance was calculated with the pump-probe delay tuned from -800 to 800 fs, while fixing the angle of incidence of the probe pulse at zero exciton-photon detuning at $33.5^\circ$ (Fig. \ref{fig:Bleaching}(c)).
Depending on the pump-probe delay, the polariton dynamics can be broadly divided into three stages, as indicated by the dashed black lines. 
When the probe pulse arrives before the pump pulse, the transient reflectance is characterized by interference fringes arising from the perturbed free induction decay (PFID) \cite{Rodek2021, feldman2026ultrafast}.
When the pump and probe pulses arrive nearly simultaneously, both upper and lower branches of the polariton dispersion blue shift due to the optical Stark effect (OSE) \cite{lamountain2021valley, zhou2024cavity}.
When the probe pulse arrives after the pump pulse, the exciton-photon coupling strength is reduced as a result of exciton bleaching, where both the upper and lower branches shift toward the exciton wavelength \cite{zhao2023exciton, zhao2025sub}.

The mechanism of the transient reflectance is further explained in the schematics in Fig. \ref{fig:Bleaching}(d) (see Supplemental Material section S3 C for generation of schematics).
According to the optical Bloch equations, exciton dynamics can be understood by the evolution of the Bloch vector $\vec{M}=\left(2\mathrm{Re}(\tilde{p}), 2\mathrm{Im}(\tilde{p}), -1+2n\right)^T$ on the Bloch sphere, with $\tilde{p}$ and $n$ representing the polarization in the rotating frame of $\omega_{\mathrm{pu}}$ and the population, respectively. 
The Bloch vector precesses under the effective magnetic field $\vec{B}=\left(\mathrm{Re}(\tilde{O}), \mathrm{Im}(\tilde{O}), \omega_{\mathrm{pu}}-\omega_{\mathrm{ex}}\right)^T$, where $\tilde{O}$ denotes the optical fields in the rotating frame of $\omega_{\mathrm{pu}}$ (Supplementary Material section S3 B). 
In the plot, the red and blue parts of the trajectory represent the evolution caused by the pump and probe pulses, respectively.
Although both pulses are of 100 fs width, the probe pulse acts as an impulse that directly perturbs the exciton, whereas the pump pulse slowly injects into the material through the high quality factor polariton mode. 
During the PFID stage, $\vec{M}$ is first perturbed by the kick of the probe pulse and precesses around the $z$ axis before its decay. 
The arrival of the strong pump pulse then drags $\vec{B}$ away from the $z$ axis nearly adiabatically while increasing its amplitude, and hence increases the precession frequency of $\vec{M}$. 
The polarizations in these two processes thus involve different frequency components, and their coherent superpositions result in interference fringes in the transient reflectance. 
The OSE stage, in contrast, is dominated by the second process as both pulses arrive simultaneously, which therefore manifests a blueshift of the exciton energy. 
For the bleaching stage, in which the probe pulse arrives well after the pump pulse, $\vec{M}$ undergoes both precession and decoherence of $\tilde{p}$ towards the $z$ axis, resulting in a residue of $n$ due to its longer relaxation time. 
This $n$ subsequently decreases the exciton-photon coupling strength in the microcavity. 

The transient polariton dispersion is further investigated as a function of the pump field amplitude at a fixed delay at  +200 fs, as shown in Fig. \ref{fig:Bleaching}(e).
When the pump field is weak as $0.1\times 10^8$ V/m, the dispersion is consistent with the linear reflectance (gray dashed lines), indicating a low level of excitation. 
As the pump field amplitude increases to $1.6$ and $2.7\times 10^8$ V/m, Rabi splitting gradually disappears due to exciton bleaching, with both polariton branches shifting toward the exciton energy.
Note that the exciton energy is lightly blue shifted due to the residue of OSE. 
These results are consistent with experimental results reported in similar materials and structures, which validates our approach \cite{zhao2023exciton, feldman2026ultrafast}. 
This example illustrates that our algorithm is capable for simulations that simultaneously involve non-perturbative nonlinear dynamics and complex linear structures.

To conclude, a rigorous framework based on TMM and the iterative Green's function method is developed and its applications in typical nonlinear photonic architectures have been demonstrated. 
Our frequency domain approach naturally relaxes the $c\Delta t < \Delta x$ constraints of conventional time-domain methods.
This feature makes it feasible to simulate ultra-broadband nonlinear processes such as octave-spanning frequency conversions in wavelength-scale nanophotonic structures.
The approach can be readily extended to periodic in-plane structures by incorporating rigorous coupled-wave analysis, providing simulation toolboxes to nonlinear photonic crystals, metasurfaces, and structured light fields. 

\textit{Acknowledgments.}- This work was supported by the National Key R\&D Program of China (grant number 2022YFA1206700). 

\textit{Data availability.}- The code used in this paper is available from the corresponding author upon request.

\bibliography{apssamp}

\onecolumngrid
{\centering
\vspace*{2em}
\textbf{\Large End Matter}\par
\vspace{2em}
}
\twocolumngrid

\renewcommand{\theequation}{A\arabic{equation}}  
\renewcommand{\thetable}{A\arabic{table}} 
\setcounter{equation}{0}  
\setcounter{table}{0}  

\appendix

\section*{\label{sec:theory} Appendix A. Detailed Description of the Unified Algorithm}

\subsection{Formulation of the Nonlinear Boundary Value Problem}
We consider the solution of Maxwell's equations in a one-dimensional (1D) multi-layer optical structure under external excitations. 
The structure contains arbitrarily distributed linear and nonlinear layers, as shown in Fig. 1(a) in the main text. 
In nonlinear layers, the nonlinear polarization $\vec{P}_{\mathrm{NL}}$ is considered in addition to the linear optical refractive index $n$
\begin{equation}
    \vec{D} = \varepsilon n^2\vec{E}+\vec{P}_{\mathrm{NL}},
\end{equation}
where $\vec{D}$ is the electric displacement vector, and $\vec{E}$ is the electric field.
The nonlinear polarization $\vec{P}_{\mathrm{NL}}$ is assumed to be a function of the local electric field
\begin{equation}
    \vec{P}_{\mathrm{NL}}(\vec{r}, t) = \vec{P}_{\mathrm{NL}}(\vec{r}, \vec{E}(\vec{r}, t)),
\end{equation}
where $\vec{r}$ is the position vector.

For a 1D multilayer system, as the system exhibits translational invariance in both temporal and transverse spatial dimensions, the total electro-magnetic (EM) field can be expanded by frequency $\omega$ and transverse momentum $\vec{k}_t$ (assuming that the interfaces between layers are perpendicular to the z direction)
\begin{equation}
    \begin{aligned}
        \vec{E}(\vec{\rho}, z, t) &= \sum_{\vec{k}_t, \omega} \mathrm{e}^{-\mathrm{i}(\vec{k}_t \cdot \vec{\rho} - \omega t)}\vec{E}(\vec{k}_t, z, \omega). \\
        \vec{H}(\vec{\rho}, z, t) &= \sum_{\vec{k}_t, \omega} \mathrm{e}^{-\mathrm{i}(\vec{k}_t \vec{\rho} - \omega t)}\vec{H}(\vec{k}_t, z, \omega),
    \end{aligned}
\end{equation}
where $H$ is the magnetic field, $\vec{\rho} = x\hat{e}_x + y\hat{e}_y$ is the transverse position vector.
$\hat{e}_{i}$ is the unit vector in the $i$-direction, with $i = x, y, z$.
Each component can be further divided into TE and TM polarizations according to the relation between the polarization directions of $\vec{E}$ and $\vec{k}_t$.
\begin{equation}
    \begin{aligned}
        \vec{E}(\vec{k_t}, z, \omega) &= E_v\hat{e}_v+(E_t\hat{e}_t+E_z\hat{e}_z), \\
        \vec{H}(\vec{k_t}, z, \omega) &= (H_t\hat{e}_t+H_z\hat{e}_z) + H_v\hat{e}_v,
    \end{aligned}
\end{equation}
where $\hat{e}_t = \vec{k}_t/|k_t|$, $\hat{e}_v = \hat{e}_t\times\hat{e}_z$ is the unit vector perpendicular to the incident plane.
The combination $(E_v, H_t, H_z)$ is defined as the TE polarization, and $(H_v, E_t, E_z)$ is defined as the TM polarization. 
The nonlinear polarization can also be decomposed into these two polarizations
\begin{equation}
    \vec{P}_{\mathrm{NL}}(\vec{k}_t, z, \omega) = P_{\mathrm{NL}, v}\hat{e}_v + (P_{\mathrm{NL}, t}\hat{e}_v + P_{\mathrm{NL}, z}\hat{e}_z).
\end{equation}
For TE polarization, the wave equation for the EM field can be written based on the electric field $E$
\begin{equation}
    \left(\frac{\partial}{\partial z^2} + \frac{n^2\omega^2}{c^2}-k_t^2\right)E_v=-\mu_0\omega^2 P_{\mathrm{NL}, v},
\end{equation}
where $\mu_0$ is the vacuum permeability.
The other two components can be related to $E_v$ through $\nabla\times\vec{E}=-\mu_0\partial_t\vec{H}$
\begin{equation}
    H_t = \frac{\mathrm{i}}{\mu_0\omega}\frac{\partial E_v}{\partial z},\quad H_z = \frac{\mathrm{i}}{\mu_0\omega}\mathrm{i}k_tE_v.
\end{equation}
At the interface ($z=d$), the continuity of the transverse field components $E_v, H_t$ gives the boundary conditions (as Eq. 1b in the main text)
\begin{equation}
    E_v|_{z=d_-}=E_v|_{z=d_+}, \quad \left.\frac{\partial E_v}{\partial z}\right|_{z=d_-}=\left.\frac{\partial E_v}{\partial z}\right|_{z=d_+}.
\end{equation}

For TM polarization, the wave equation for the EM field can be written based on the magnetic field $H$
\begin{equation}
    \left(\frac{\partial}{\partial z^2} + \frac{n^2\omega^2}{c^2}-k_t^2\right)H_v = -\mathrm{i}\omega Q_{\mathrm{NL, v}},
\end{equation}
where $Q_{\mathrm{NL, v}}=(\nabla\times\vec{P}_{\mathrm{NL}})_v$, and the other two components can be related to $H_v$ through $\nabla\times\vec{H} = \partial_t\vec{D}$
\begin{equation}
    E_t = -\frac{\mathrm{i}}{\varepsilon_0n^2\omega}\frac{\partial H_v}{\partial z}-\frac{P_{\mathrm{NL}, t}}{\varepsilon_0 n^2}, \quad E_z = \frac{-\mathrm{i}}{\varepsilon_0 n^2 \omega}\mathrm{i}k_tH_v -\frac{P_{\mathrm{NL}, z}}{\varepsilon_0 n^2},
\end{equation}
where $\varepsilon_0$ is the vacuum permittivity.
At the interface ($z=d$), the continuity of the transverse field components $E_t, H_v$ gives the boundary conditions
\begin{equation}
\begin{aligned}
    H_v|_{z=d_-}&=H_v|_{z=d_+}, \\
    \left.\frac{\mathrm{i}}{\omega}\frac{\partial H_v}{\partial z}\right|_{z=d_-}+P_{\mathrm{NL},t}|_{z=d_-}&=\left.\frac{\mathrm{i}}{\omega}\frac{\partial H_v}{\partial z}\right|_{z=d_+} + P_{\mathrm{NL},t}|_{z=d_+}.
\end{aligned}
\end{equation}

Considering the total number of nonlinear layers in the structure is $N$, and the interval of the $n$ th nonlinear layer is $[a_n, b_n]$, the domain of the above wave equations can be denoted as
\begin{equation}
    z\in[a_1, b_1], [a_2, b_2], \dots, [a_N, b_N].
\end{equation}

The wave equations written above should be solved with appropriate boundary conditions. In the linear case, the transfer matrix method can be used to solve the EM field of a one-dimensional multilayer structure conveniently, as it bridges the tangential field components on both sides of the linear layer with a matrix. Here, we implement the transfer matrix method to build up the boundary conditions. Define the vector of tangential field components for the TE polarization as
\begin{equation}
    V_{\mathrm{TE}}(z)=\begin{pmatrix}
        E_{v}(z) \\ \eta_0 H_{t}(z)
    \end{pmatrix} = \begin{pmatrix}
        E_v(z) \\ \frac{\mathrm{i}c}{\omega}\frac{\partial E_v(z)}{\partial z}
    \end{pmatrix} = S(z),
\end{equation}
where $\eta_0$ is the vacuum impedance.
$S$ is homogeneous with respect to $E_v$. 
For the TM polarization, we can define that
\begin{equation}
    \begin{aligned}
        V_{\mathrm{TM}}(z) &= \begin{pmatrix}
            E_t(z) \\ \eta_0H_v(z)
        \end{pmatrix} = U(z) + W(z), \\
        U(z) &= \begin{pmatrix}
            -\frac{\mathrm{i}}{\varepsilon_0n^2\omega}\frac{\partial H_v(z)}{\partial z} \\ \eta_0 H_v(z)
        \end{pmatrix},  \\
        W(z) &= \begin{pmatrix}
            -\frac{P_{\mathrm{NL}, t}(z)}{\varepsilon_0n^2} \\ 0
        \end{pmatrix}. \\
    \end{aligned}
\end{equation}
Here, $U$ is the homogeneous tangential field with respect to $H_v$, while $W$ is the residual inhomogeneous tangential field.
The difference between $V_{\mathrm{TE}}$ and $V_{\mathrm{TM}}$ originates from the asymmetry between nonlinear electric and magnetic polarization.
In linear layers, the vector can also be expressed with the electric field amplitude of the forward ($E_f$) and backward ($E_b$) propagating wave 
\begin{equation}
    V_{\mathrm{TE/TM}}=\begin{pmatrix}
        E_f + E_b \\ \gamma_{\mathrm{TE/TM}}(E_f-E_b)
    \end{pmatrix},
\end{equation}
where $\gamma_{\mathrm{TE}}=n\cos\theta$, $\gamma_{\mathrm{TM}}=n/\cos\theta$, and $\theta$ is the angle of propagation with respect to $\hat{e}_z$ in the corresponding layer.
For a linear layer spanning $z\in[z_1, z_2]$, the transfer matrix linking the tangential field vector can be defined as
\begin{equation}
    \hat{M}_{\text{TE/TM}}(z_1, z_2) = \begin{pmatrix}
        \cos\delta & \frac{\mathrm{i}\sin\delta}{\gamma_{\text{TE/TM}}} \\
        \mathrm{i}\gamma_{\text{TE/TM}}\sin\delta & \cos\delta
    \end{pmatrix},
\end{equation}
\begin{equation}
    V_{\text{TE/TM}}(z_1)=\hat{M}_{\text{TE/TM}}(z_1, z_2)V_{\text{TE/TM}}(z_2),
\end{equation}
where $\delta=\omega n d\cos\theta/c$, $d=z_2-z_1$.
For multiple layers, one can multiply the transfer matrix of each layer sequentially to obtain the total transfer matrix of the structure.

The boundary conditions between two nonlinear layers can be expressed as
\begin{equation}
    V(b_{n-1}) = \hat{M}_{n-1, n}V(a_n),
\end{equation}
where $\hat{M}_{n-1, n}$ is the transfer matrix of the linear layers between the $n-1$ th and the $n$ th nonlinear layers. 
The boundary conditions at the top interface of the structure can be expressed as
\begin{equation}
  \begin{pmatrix}
      E_{\mathrm{inc}} + E_{\mathrm{ref}} \\
      \gamma(E_{\mathrm{inc}} - E_{\mathrm{ref}})
  \end{pmatrix} = V_{\mathrm{top}} = \hat{M}_{\mathrm{top}}V(a_1),
\end{equation}
where $E_{\mathrm{inc}}$ and $E_{\mathrm{ref}}$ are the amplitude of incidence and reflection, respectively. The boundary conditions at the bottom interface of the structure can be expressed as
\begin{equation}
    \hat{M}_{\mathrm{bot}}^{-1}V(b_N) = V_{\mathrm{bot}}=\begin{pmatrix}
        E_{\mathrm{trans}} \\ \gamma E_{\mathrm{trans}}
    \end{pmatrix},
\end{equation}
where $E_{\mathrm{trans}}$ is the amplitude of transmission.

So far, we have formulated the complete problem as a nonlinear boundary value problem (BVP).

\subsection{Solution Decomposition According to BVP Inhomogeneity}
To solve this problem, we decompose the total solution into linear and nonlinear parts
\begin{equation}
    \begin{aligned}
        E_{v} &= E_{v, \mathrm{L}} + E_{v, \mathrm{NL}}, \\
        V_{\mathrm{TE}}(z) &= S_{\mathrm{L}}(z) + S_{\mathrm{NL}}(z),
    \end{aligned}
\end{equation}
\begin{equation}
    \begin{aligned}
        H_{v} &= E_{v, \mathrm{L}} + E_{v, \mathrm{NL}}, \\
        V_{\mathrm{TM}}(z) &= U_{\mathrm{L}}(z) + U_{\mathrm{NL}}(z) + W(z),
    \end{aligned}
\end{equation}
The linear part of the field satisfies linear wave equation and boundary conditions with external incidence.
For the TE polarization,
\begin{equation}
    \begin{aligned}
        \left(\frac{\partial}{\partial z^2} + \frac{n^2\omega^2}{c^2}-k_t^2\right)E_{v, \mathrm{L}} &= 0, \\
       \hat{M}_{\mathrm{top}}S_{\mathrm{L}}(a_1) &=  \begin{pmatrix}
      E_{\mathrm{inc}} + E_{\mathrm{ref, L}} \\
      \gamma_{\mathrm{TE}} (E_{\mathrm{inc}} - E_{\mathrm{ref, L}})
  \end{pmatrix},\\
  \hat{M}_{n-1, n}S_{\mathrm{L}}(a_n) &= S_{\mathrm{L}}(b_{n-1}), \\
  \hat{M}_{\mathrm{bot}}^{-1}S_{\mathrm{L}}(b_N) &= \begin{pmatrix}
        E_{\mathrm{trans, L}} \\ \gamma_{\mathrm{TE}} E_{\mathrm{trans, L}}
    \end{pmatrix}.
    \end{aligned}
\end{equation}
For the TM polarization
\begin{equation}
    \begin{aligned}
        \left(\frac{\partial}{\partial z^2} + \frac{n^2\omega^2}{c^2}-k_t^2\right)H_{v, \mathrm{L}} &= 0, \\
        \hat{M}_{\mathrm{top}}U_{\mathrm{L}}(a_1) &= \begin{pmatrix}
      E_{\mathrm{inc}} + E_{\mathrm{ref, L}} \\
      \gamma_{\mathrm{TM}} (E_{\mathrm{inc}} - E_{\mathrm{ref, L}})
  \end{pmatrix} ,\\
  \hat{M}_{n-1, n}U_{\mathrm{L}}(a_n) &= U_{\mathrm{L}}(b_{n-1}) , \\
  \hat{M}_{\mathrm{bot}}^{-1}U_{\mathrm{L}}(b_N) &= \begin{pmatrix}
        E_{\mathrm{trans, L}} \\ \gamma_{\mathrm{TM}}  E_{\mathrm{trans, L}}
    \end{pmatrix}.
    \end{aligned}
\end{equation}
The linear part of the field can be directly obtained by TMM.
The nonlinear part of the field satisfies nonlinear wave equation and boundary conditions without external incidence. 
For the TE polarization
\begin{equation}\label{eq:ENL}
    \begin{aligned}
        \left(\frac{\partial}{\partial z^2} + \frac{n^2\omega^2}{c^2}-k_t^2\right)E_{v, \mathrm{NL}} &= -\mu_0\omega^2P_{\mathrm{NL}, v}, \\
        \hat{M}_{\mathrm{top}}S_{\mathrm{NL}}(a_1) &= \begin{pmatrix}
      E_{\mathrm{ref, NL}} \\
      -\gamma_{\mathrm{TE}}  E_{\mathrm{ref, NL}}
  \end{pmatrix} ,\\
  \hat{M}_{n-1, n}S_{\mathrm{NL}}(a_n) &= S_{\mathrm{NL}}(b_{n-1}) , \\
  \hat{M}_{\mathrm{bot}}^{-1}S_{\mathrm{NL}}(b_N) &= \begin{pmatrix}
        E_{\mathrm{trans, NL}} \\ \gamma_{\mathrm{TE}}  E_{\mathrm{trans, NL}}
    \end{pmatrix}.
    \end{aligned}
\end{equation}
This boundary value problem with an inhomogeneous differential equation but homogeneous boundary conditions.
For the TM polarization, we can further divide $H_{v,\mathrm{NL}}$ into two parts
\begin{equation}
    H_{v,\mathrm{NL}} = H^{\mathrm{i}}_{v,\mathrm{NL}}+H^{\mathrm{h}}_{v,\mathrm{NL}},
\end{equation}
where $H^{\mathrm{i}}_{v,\mathrm{NL}}$ satisfies the linear wave equation and inhomogeneous boundary conditions without external incidence, and $H^{\mathrm{h}}_{v,\mathrm{NL}}$ satisfies the nonlinear wave equation and homogeneous boundary conditions without external incidence
\begin{equation}\label{eq:HNLi}
    \begin{aligned}
        \left(\frac{\partial}{\partial z^2} + \frac{n^2\omega^2}{c^2}-k_t^2\right)H^{\mathrm{i}}_{v, \mathrm{NL}} &=0, \\
        \hat{M}_{\mathrm{top}}(U^{\mathrm{i}}_{\mathrm{NL}}(a_1) + W(a_1)) &= \begin{pmatrix}
      E^{\mathrm{i}}_{\mathrm{ref, NL}} \\
      -\gamma_{\mathrm{TM}}  E^{\mathrm{i}}_{\mathrm{ref, NL}})
  \end{pmatrix},\\
  \hat{M}_{n-1, n}(U^{\mathrm{i}}_{\mathrm{NL}}(a_n)+W(a_n)) &=  U^{\mathrm{i}}_{\mathrm{NL}}(b_{n-1}) + W(b_{n-1}), \\
  \hat{M}_{\mathrm{bot}}^{-1}(U^{\mathrm{i}}_{\mathrm{NL}}(b_N) +W(b_N))&= \begin{pmatrix}
        E^{\mathrm{i}}_{\mathrm{trans, NL}} \\ \gamma_{\mathrm{TM}} E^{\mathrm{i}}_{\mathrm{trans, NL}}
    \end{pmatrix}.
    \end{aligned}
\end{equation}
\begin{equation}\label{eq:HNLh}
    \begin{aligned}
        \left(\frac{\partial}{\partial z^2} + \frac{n^2\omega^2}{c^2}-k_t^2\right)H^{\mathrm{h}}_{v, \mathrm{NL}} &= -\mathrm{i}\omega Q_{\mathrm{NL}, v}, \\
        \hat{M}_{\mathrm{top}}U^{\mathrm{h}}_{\mathrm{NL}}(a_1) &= \begin{pmatrix}
      E^{\mathrm{h}}_{\mathrm{ref, NL}} \\
      -\gamma_{\mathrm{TM}}  E^{\mathrm{h}}_{\mathrm{ref, NL}})
  \end{pmatrix},\\
  \hat{M}_{n-1, n}U^{\mathrm{h}}_{\mathrm{NL}}(a_n) &=  U^{\mathrm{h}}_{\mathrm{NL}}(b_{n-1}), \\
  \hat{M}_{\mathrm{bot}}^{-1}U^{\mathrm{h}}_{\mathrm{NL}}(b_N)&= \begin{pmatrix}
        E^{\mathrm{h}}_{\mathrm{trans, NL}} \\ \gamma_{\mathrm{TM}}  E^{\mathrm{h}}_{\mathrm{trans, NL}}
    \end{pmatrix}.
    \end{aligned}
\end{equation}

Our decomposition of the total solution can be understood from its one-to-one correspondence with the inhomogeneities originating from different sources in this nonlinear BVP.
In both TE and TM cases, the nonlinear polarization gives rise to the  inhomogeneous driving term in the wave equation.
However, the composition of the inhomogeneity in the boundary conditions differs between the TE and TM polarizations.
For the TE polarization, the inhomogeneity of its boundary conditions arises entirely from the external incident field.
For the TM polarization, the inhomogeneity of its boundary conditions can be divided into two parts: the external incident field and the residual tangential field.
Correspondingly, each component of our decomposition of the complete solution individually satisfies one type of these inhomogeneities.
For the TE polarization, $E_{\text{L}}$ satisfies the inhomogeneous boundary condition given by the external incident field, while $E_{\text{NL}}$ satisfies the inhomogeneous driving term in the wave equation given by the nonlinear polarization, as summarized in Tab. A1.
For the TM polarization, $H_{\text{L}}$ and $H_{\text{NL}}^h$ satisfy the same inhomogeneous condition as $E_{\text{L}}$ and $E_{\text{NL}}$, respectively, while an additional $H_{\text{NL}}^{\text{i}}$ is introduced to satisfy the inhomogeneous boundary condition induced by the residual tangential field, as summarized in Tab. A2.

\begin{table*}[ht]
\label{tab:TEDecomp}
\centering
\setlength{\tabcolsep}{12pt}
\setlength{\arrayrulewidth}{0.05em}
\renewcommand{\arraystretch}{1.5}
\caption{Decomposition of Inhomogeneity for TE Polarization}
\begin{tabular}{c c c}
\Xhline{2\arrayrulewidth}
\multirow{3}{*}{Components} & \multicolumn{2}{c}{Inhomogeneity to be satisfied} \\
\cline{2-3}
 & Boundary Conditions & Wave Equation \\
\cline{2-3}
 & $E_{\text{inc}}$ & Nonlinear Polarization $P_{\mathrm{NL}}$ \\
\hline
$E_{\text{L}}$        & $\checkmark$ & \\
$E_{\text{NL}}$       & & $\checkmark$ \\
\Xhline{2\arrayrulewidth}
\end{tabular}
\end{table*}

\begin{table*}[ht]
\label{tab:TMDecomp}
\centering
\setlength{\tabcolsep}{12pt}
\setlength{\arrayrulewidth}{0.05em}
\renewcommand{\arraystretch}{1.5}
\caption{Decomposition of Inhomogeneity for TM Polarization}
\begin{tabular}{c c c c}
\Xhline{2\arrayrulewidth}
\multirow{3}{*}{Components} & \multicolumn{3}{c}{Inhomogeneity to be satisfied} \\
\cline{2-4}
 & \multicolumn{2}{c}{Boundary Conditions} & Wave Equation \\
\cline{2-4}
 & External Incidence $E_{\text{inc}}$ & Residual Tangential Field $W$ & Nonlinear Polarization $P_{\mathrm{NL}}$ \\
\hline
$H_{\text{L}}$        & $\checkmark$ & & \\
$H_{\text{NL}}^{\text{i}}$ & & $\checkmark$ & \\
$H_{\text{NL}}^{\text{h}}$ & & & $\checkmark$ \\
\Xhline{2\arrayrulewidth}
\end{tabular}
\end{table*}

\subsection{Iterative Green's Function Method}

As the linear part of the solution ($E_{\text{L}}, H_{L})$ can be solved directly by TMM, the remaining problem is to obtain a self-consistent solution of the nonlinear part.
Here, we introduce a fixed point iteration with Green's function method.
The Green's function $G(z,s)$ for the nonlinear wave equation is defined as
 \begin{equation}\label{eq:GreenWaveEq}
   \left(\frac{\partial^2}{\partial z^2}+\frac{n^2\omega^2}{c^2}-k_t^2\right)G(z,s) =\delta(z-s),
\end{equation}
with corresponding homogeneous boundary conditions.
For $E_{\mathrm{NL}}$ and $H_{\mathrm{NL}}^{\mathrm{i}}$, Eq. \ref{eq:ENL} and \ref{eq:HNLi} can be expressed in the form of an integral
\begin{equation}\label{eq:ENLGreen}
    E_{\mathrm{NL}} = -\mu_0\omega^2 \int\mathrm{d}s\, G_E(z,s)P_{\mathrm{NL},v},
\end{equation}
\begin{equation}\label{eq:HNLhGreen}
    H^{\mathrm{h}}_{\mathrm{NL}} = -\mathrm{i}\omega \int\mathrm{d}s\, G_H(z,s)Q_{\mathrm{NL},v},
\end{equation}
where the subscripts $E$ and $H$ indicate that $G$ is the Green's function for the electric field and the magnetic field, respectively.
In addition, for the TM polarization, the boundary induced term $H_{\mathrm{NL}}^{\mathrm{i}}$ needs to be obtained through Eq. \ref{eq:HNLi}, which can be solved by assuming a superposition of general solutions of the wave equation and extracting the coefficients with boundary conditions.

To solve Eq. \ref{eq:ENLGreen}, \ref{eq:HNLhGreen} and \ref{eq:HNLi} self-consistently, we adopt the following iteration
\begin{enumerate}
    \item Solve the linear part of the EM field with TMM, obtaining $E_{v, \mathrm{L}}$, $H_{v, \mathrm{L}}$;
    \item Initially set $P^{(0)}_{\mathrm{NL}, v}=0$, $Q_{\mathrm{NL}, v}^{(0)}=0$;
    \item Begin fixed point iteration, set $l=1$
    \begin{enumerate}
        \item In the $l$ th iteration, treat $\vec{P}^{(l-1)}_{\mathrm{NL}}$ as constant, directly integrate Eq. \ref{eq:ENLGreen}, \ref{eq:HNLhGreen} to obtain $E_{v, \mathrm{NL}}^{(l)}$ and $H_{v, \mathrm{NL}}^{\mathrm{h}, (l)}$, solve Eq. \ref{eq:HNLi} to obtain $H_{v, \mathrm{NL}}^{\mathrm{i}, (l)}$;
        \item Calculate $\vec{P}^{(l)}_{\mathrm{NL}}$ with 
        \begin{equation}
            \vec{P}^{(l)}_{\mathrm{NL}} = \vec{P}_{\mathrm{NL}}\left(E_{v,\mathrm{L}}+E^{(l)}_{v,\mathrm{NL}}, H_{v,\mathrm{L}}+H^{\mathrm{h},(l)}_{v,\mathrm{NL}}+H^{\mathrm{i},(l)}_{v,\mathrm{NL}}\right);
        \end{equation}
        \item Calculate the error between the two sides of the Eq. \ref{eq:ENL} and \ref{eq:HNLh}. 
        If the error is below tolerance, stop iteration.
        If not, set $l = l+1$.
    \end{enumerate}
    \item Output $E_v = E_{v,\mathrm{L}}+E_{v,\mathrm{NL}}$, $H_v = H_{v,\mathrm{L}}+H_{v,\mathrm{NL}}^{\mathrm{h}}+H_{v,\mathrm{NL}}^{\mathrm{i}}$.
    
\end{enumerate}

\end{document}